\documentstyle[12pt]{article}

\def\be{\begin{equation}}
\def\ee{\end{equation}}
\def\ba{\begin{array}{c}}
\def\ea{\end{array}}
\def\p{\partial}
\def\ben{$$}
\def\een{$$}
\begin{document}

\titlepage
\vspace*{4cm}

 \begin{center}
{\Large \bf
${\cal PT}-$symmetric square well
 }

\end{center}

\vspace{5mm}

 \begin{center}

Miloslav Znojil
\vspace{3mm}

\'{U}stav jadern\'e fyziky AV \v{C}R, 250 68 \v{R}e\v{z},
Czech Republic\\

e-mail: znojil@ujf.cas.cz

\end{center}

\vspace{5mm}

\section*{Abstract}

Below a (comparatively large) measure of non-Hermiticity
$Z=Z_0^{(crit)}>0$ of a ${\cal PT}$ symmetrically complexified
square well, bound states are constructed non-numerically.  All
their energies prove real and continuous in the (Hermitian) limit
$Z \to 0$.  Beyond the threshold $Z_0^{(crit)}$ (and, in general,
beyond $Z_m^{(crit)}$ at $m=0, 1,\ldots$) the lowest two real
energies (i.e., $E_{2m}$ and $E_{2m+1}$) are shown to merge and
disappear.

\vspace{9mm}

\noindent
PACS 03.65.Ge, 03.65.Fd



\newpage

\section{Introduction}

An interest of physicists in complex potentials $V(x)= {\rm
Re}\,V(x) + i\,{\rm Im}\,V(x)$ with the generalized symmetry
property
 \be
{\rm Re}\,V(x) ={\rm Re}\,V(-x),
\ \ \ \ \ \ \ \
{\rm Im}\,V(x) =-{\rm Im}\,V(-x)
\label{PTs}
 \ee
dates back to the perturbative study of imaginary cubic anharmonic
oscillators
 \be
V(x) = \omega\,x^2 + i\,\lambda\,x^3
\label{ixnatri}
 \ee
by Caliceti et al \cite{Caliceti}. In the early nineties, an
increase of this interest \cite{Alvarez} was inspired by the role
of the imaginary cubic anharmonicity in field theory where eq.
(\ref{PTs}) mimics the fundamental parity times time-reversal (=
${\cal PT}$) symmetry of phenomenological Lagrangians \cite{BM}.
Under certain circumstances, the ${\cal PT}$ symmetric and
Hermitian models can even prove mathematically equivalent
\cite{BG}.

In the late nineties, Bender and Boettcher \cite{BB} analysed
the one-parametric family of the power-law $\omega \to 0$ models
 \be
V(x) = x^2\,( i\,x)^\delta
\label{ixnadelta}
 \ee
by the quasi-classical and purely numerical means.  On this basis
they conjectured that all the $ \delta > 0$ bound-state energies
$E_n=E_n(\delta)$ form a real and discrete, smooth continuation of
the well known harmonic-oscillator $\delta=0$ spectrum.  An
extension of this study inspired them later to apply the
conjecture (formulated, originally, by Bessis at $\delta=1$
\cite{Bessis}) to a still broader class of interactions.  Within
the resulting ``generalized" (so called ${\cal PT}$ symmetric)
quantum mechanics \cite{BBM}, there appears a growing number of
interesting studies, promoting the ideas of supersymmetry
\cite{Junker,Ioffe}, exact semiclassical techniques \cite{Pham},
functional analysis \cite{Mez} and perturbation theory \cite{ix3}.
All of them provide a consistent picture of a theory with certain
(not quite well understood) limitations. Even in the above
``guiding" example (\ref{ixnadelta}) the ${\cal PT}$ symmetry
breaks down spontaneously at $\delta < 0$ \cite{BB}.

Phenomenological appeal of the ${\cal PT}$ symmetric modifications
need not even stop before the ``sacred" quantum electrodynamics
\cite{QED}.  Still, the formalism abounds in open questions and
its mathematical foundations are mostly conjectures.  This is the
reason why the attention quickly spreads from the
phenomenologically oriented eq. (\ref{ixnadelta}) to its exactly
solvable alternatives.  In this direction, the partially solvable
extension of the $\delta=2$ quartic case \cite{BBjpa} and the
discovery of the exact solvability of the certain version of the
$\delta \to \infty$ limiting case \cite{Junker} were the first and
encouraging successes. They were followed by the ${\cal PT}$
symmetric regularization of the more-dimensional harmonic
oscillator \cite{HOPT} and by the formulation and solution of many
further shape-invariant models \cite{RM}-\cite{shapelist}.

Once we move beyond the domain of analytic potentials, numerical
studies provide significantly less encouraging results
 \cite{BBM}. This is the reason why the ``most elementary" square
well
 \be
\ba
{\rm Re}\,V(x) =0, \ \ \ \ \ x \in (-1,1)\\
{\rm Im}\,V(x) =Z, \ \ \ \ \ x \in (-1,0)\\
{\rm Im}\,V(x) =-Z, \ \ \ \ \ x \in (0,1)
\ea
\label{SQW}
 \ee
has always been ${\cal PT}$ symmetrized just in the (various,
non-equivalent) $\delta \to \infty$ limits of
eq.~(\ref{ixnadelta})~\cite{Junker,my4,BSQW}.

\section{Solution}

Presumably, the reason for absence of the ``forgotten" ${\cal PT}$
symmetric square well (\ref{SQW}) in the literature lies in an
ambiguity of its continuation beyond the discontinuities at $x =
\pm 1$.  We are going to treat this problem simply by imposing the
Dirichlet boundary conditions on our (complex) wave functions,
 \be
\psi(\pm 1) = 0.
\label{bc}
 \ee
Having made the latter decision the explicit computations are
really elementary. Their essence lies in the easy construction
of the general right and left solutions
 \be
\ba \psi_{+} = c_{+} e^{\kappa\,x} +d_{+} e^{-\kappa\,x}\\
\psi_{{-}} = c_{-} e^{\kappa^{\ast}\,x}
 +d_{-} e^{-\kappa^{\ast}\,x}
\ea
\label{gens}
 \ee
with the complex $\kappa$ and its conjugate $\kappa^\ast$. We
re-parametrize the (by assumption, real) energies $E = t^2-s^2$
and the measure of non-Hermiticity $Z = 2st$.  This gives the
complex exponents $\kappa=s-it$ since $\kappa^2=V(x)-E$ is also
complex (and constant for all $x \in (0,1)$).  Our solutions
(\ref{gens}) are made explicit, for every real $E$ and $Z$, when
we use the inverse formulae
 \ben
t=\frac{1}{\sqrt{2}}\,\left ({E+\sqrt{E^2+Z^2}}\right )^{1/2},
 \een
 \ben
s=\frac{Z}{\sqrt{2}}\,\left ({E+\sqrt{E^2+Z^2}}\right )^{-1/2}.
 \een
In the light of the ${\cal PT}$ symmetry of our Hamiltonian
$H={\cal PT} H {\cal PT}$, the product ${\cal PT}\psi(x) \equiv
\psi^\ast(-x)$ will satisfy the same Schr\"{o}dinger equation as
$\psi(x)$.  Hence, in the origin, we are permitted to normalize
our bound states in a ${\cal PT}$ symmetric way,
 \ben
\psi_{+}(0)=\psi_{-}(0)=1, \ \ \ \ \
\p_x\psi_{+}(0)=\p_x\psi_{-}(0)=i\,A.
 \een
These conditions contain a free real parameter $A$ and are
equivalent to the matching of wave functions,
 \ben
\ba c_{\pm} = 1 - d_{\pm}, \\ \kappa\,(1-2\,d_{+}) = \kappa^\ast
(1-2\,d_{-})=iA. \ea
 \een
This implies that we know all the coefficients in eq.
(\ref{gens}),
 \ben
d_{+}=\frac{1}{2} - i\,\frac{A}{2\kappa}, \ \ \ \ \ \
d_{-}=\frac{1}{2} - i\,\frac{A}{2\kappa^\ast}.
 \een
It is easy to satisfy the external boundary conditions
(\ref{bc}) and reduce them to the elementary prescription
 \be
i\,A=-\kappa\,\coth \kappa.
\label{bcexpl}
 \ee
In terms of the real parameters $s$ and $t$ this represents
a system of two algebraic equations,
 \be
A=-\frac{s}{\tan t} + t\, \tanh s = s\, \tan t +
\frac{t}{\tanh s}.
\label{graph}
 \ee
Its first part defines the (necessarily, real) value of $A=A(s,t)$.
A re-arrangement
of the second algebraic equation is elementary and gives the rule
 \be
-2t\,\sin 2t = 2s\,\sinh 2s.
\label{hical}
 \ee
As long as both its sides depend on the mere absolute values of
the respective variables $t$ and $s$, we may pick up $t \geq 0$,
insert the definition of $s=Z/2t$ and solve this equation
numerically.

\section{Discussion}

By construction, the spectrum of the real energies $E=E_n$ is
defined in terms of the roots $t_n$ of eq. (\ref{hical}),
$E_n=t^2_n -s^2_n$, $s_n=Z/2t_n$, $n = 0, 1, \ldots$.  After we
re-scale $t \to T = 2t/\pi$, we immediately see that in the
limit $Z \to 0$ this spectrum degenerates to the known Hermitian
one, with $T_n=n+1$ etc.  From our equation (\ref{hical}) it is
obvious that at the very small $Z$ (and, hence, $s \approx
0$), the change of the above roots $T_n$ (and, of course,
energies) remains very small as well.

At the larger (and, say, positive) $Z$, the analysis of our
quantization condition (\ref{hical}) becomes significantly
simplified by the re-scalings $s \to S_0 = 2\,s\,\sinh 2s$ and
$S_0 \to S$ such that, say, $S_0=4\,\sinh^2 S$.  Both these steps
represent a one-to-one mapping exhibiting the strict leading-order
quasi-linearity $s \sim S$ achieved at both ends of our half-axis,
i.e., for $s \approx 0$ as well as for $s \to \infty$.  Moreover,
this introduces just a minimal deformation of the scale of the
coordinate $s$ (practically invisible on a picture) and replaces
equation (\ref{hical}) by a new one,
 \be
-\pi\,T\,\sin \pi T = 4\,\sinh^2 S.
\label{hicalbe}
 \ee
This new relation is explicitly solvable. The resulting analytic and
$Z-$independent formula $S=X(T)$ defines a
 curve in our new $S-T$ plane.  Its $T <
14$ part is displayed in Figure~1.

In order to determine the separate roots $T_n$ (and, hence, the
spectrum of energies), it remains for us to recollect the
$Z-$dependent, hyperbolic constraint $s=Z/2t$. After we translate
its form in our new variables, $S = Y(Z,T)$, we discover that it
has a monotonous, hyperbolic shape which depends on $Z$. Figure~1
offers a few samples. We may conclude that each point of the
intersection of our two curves $X(T)$ and $Y(Z,T)$ determines a
$Z-$dependent root $T=T(Z)$ and, hence, a real energy.

The $Z-$ dependence of the roots $T(Z)$ is, in general, smooth.  A
non-perturbative effect is only encountered at certain critical
values $Z=Z^{(crit)}$. In the vicinity of these points, the two
leftmost roots $T_0(Z) < T_1(Z)$, $Z < Z^{(crit)}$ merge in a
single, doubly degenerate real root $T_0(Z^{(crit)}) =
T_1(Z^{(crit)})$. Figure~1 gives two examples and shows that even
the smallest one of these critical values is already quite large
($Z^{(crit)}_0 \approx 4.48$). Below this bound we may summarize
that

 \begin{itemize}

\item
our complexified square well
 generates the infinite set of  real
energies;

\item
the roots $T_n$ which define these energies
are almost equidistant, especially at the higher $n$;

\item
as expected, the standard square-well-type behaviour of the
spectrum
 is reproduced at the small $Z$ and for the
highly excited states.

\end{itemize}

 \noindent
In the strongly non-Hermitian domain, the lowest doublets of
states subsequently disappear.  Presumably, their energies
dissolve in conjugate pairs in complex plane. The ${\cal PT}$
symmetry of their wave functions breaks down.  The related
solutions become ``unphysical" and have to be omitted in a way
paralleling the similar ``disappearance of states" at $\delta<0$
in the model (\ref{ixnadelta}) of ref. \cite{BB}. The resulting
sudden upward jump of the ground-state energy definitely enters
the list of paradoxes, emerging in the other exactly solvable
models. Their present list already involves the unavoided level
crossings in the harmonic and Coulomb oscillators
\cite{HOPT,Coulomb}, the anomalously large excitations in the
${\cal PT}$ symmetrized but bounded Rosen-Morse field \cite{RM},
some unexpected manifestations of the strong singularities in the
P\"{o}sch-Teller and Eckart models \cite{Eckart}, and a
spontaneous re-ordering of levels in the Morse asymmetric
well~\cite{Morse}.

In this context, the role of the complexified square well is as
exceptional as in the Hermitian limit.  We may expect that its
generalizations with more points of discontinuity will remain
tractable analytically.  This could be of a significant help, say,
within perturbation theory \cite{matchingpt}.  In the more
pragmatic numerical setting, one should recollect that the so
called Pr\"{u}fer transformation \cite{Pruefer} (i.e.,  nothing
but a square-well-inspired use of the {\em locally} exponential
solutions)  found a firm place in the standard computer software
\cite{Maple}.  Last but not least, one has to keep in mind that in
the Hermitian quantum mechanics the use of the locally constant
forces could also clarify the various manifestations of the
pertaining Sturm Liouville theory \cite{Ince}.  An appropriate
${\cal PT}$ symmetrization of this theory is expected to be quite
a difficult task \cite{BBSturm}. In such a direction, also the
knowledge of our present solutions could mediate a further
progress, hopefully, in the near future.

\section*{Acknowledgement}

Partially supported by the GA AS grant Nr. A 104 8004.

\end{document}